# Single-photon Signal from Neutralinos at LEP2

S. Ambrosanio [a, 1], B. Mele [b], G. Montagna [c],
O. Nicrosini [d, 2] and F. Piccinini [e].

[a] University of Michigan, Ann Arbor, MI, USA
[b] INFN, Sezione di Roma 1, Italy
[c] Dipartimento di Fisica Nucleare e Teorica and INFN, Sezione di Pavia, Italy
[d] CERN, TH Division, Geneva, Switzerland
[e] INFN, Sezione di Pavia, Italy

## Abstract

The production of invisible pairs of lightest neutralinos accompanied by a large-angle hard photon in the reaction $e^+e^- \to \tilde{\chi}_1^0 \tilde{\chi}_1^0 \gamma$ is studied at LEP2 energies. The most general gaugino/higgsino composition of the $\tilde{\chi}_1^0$ within the Minimal Supersymmetric Standard Model is assumed. The spectrum of the observed photon is derived within the framework of the $p_t$-dependent structure-function approach, whose accuracy is assessed to be within the foreseen experimental accuracy at LEP2. Higher-order QED corrections due to undetected initial-state radiation are also included. A comparison with the Standard Model main background from $e^+e^- \to \nu\bar{\nu}\gamma$ is performed for optimized photon kinematical cuts. Quantitative conclusions on the signal/background ratio are given for a wide range of values of the SUSY parameters.

e-mail:
ambros@umich.edu, mele@roma1.infn.it, montagna@pv.infn.it,
nicrosini@vxcern.cern.ch, piccinini@pv.infn.it

---

[1] Supported by a post-doctoral INFN fellowship, Italy.
[2] Permanent address: INFN, Sezione di Pavia, Italy.

# 1  Introduction

In the Minimal Supersymmetric Standard Model (MSSM) with conserved R parity [1], the lightest supersymmetric (SUSY) partner of the known particles is predicted to be stable, neutral and weakly interacting. The usual candidate for this role is the lightest of the four neutralinos. When produced in a normal collider experiment, the lightest neutralino ($\tilde{\chi}_1^0$) is expected to generate missing energy and momentum in the final state. Furthermore, SUSY partners must be produced in pairs. In $e^+e^-$ collisions, then, from the point of view of the available phase-space, the $\tilde{\chi}_1^0$-pair production would be the easiest SUSY channel. Actually, the occurrence of this invisible channel could only be detected by the measurement of the $Z$ invisible width (if $m_{\tilde{\chi}_1^0} < M_Z/2$). However, the coupling $Z\tilde{\chi}_1^0\tilde{\chi}_1^0$ turns generally out to be small and the presence of a sizeable neutralino contribution to $\Gamma_Z^{inv}$ can only rarely exclude regions of the parameter space not already ruled out by searches for visibile channels. Hence, in $e^+e^-$ collisions, the best way to produce, at tree level, observable final states from neutralinos is through the channel $e^+e^- \to \tilde{\chi}_1^0\tilde{\chi}_2^0$, where $\tilde{\chi}_2^0$ is the next-to-lightest neutralino [2]. This process can be competitive with the production of light-chargino pairs for particular regions of the SUSY parameter space [3]. In this work, we study instead the production of $\tilde{\chi}_1^0$ pairs accompanied by a hard large-angle photon

$$e^+e^- \to \tilde{\chi}_1^0\tilde{\chi}_1^0\gamma$$

at the CERN LEP2, assuming as a typical c.m. energy $\sqrt{s} \simeq 190$ GeV. The corresponding Feynman graphs are shown in Fig. 1. Its experimental signature consists of single-photon events with missing energy and momentum. For moderate cuts on the photon energy, the available phase space for this process is larger than that for the $\tilde{\chi}_1^0\tilde{\chi}_2^0$ final state. On the other hand, the radiative $e^+e^- \to \tilde{\chi}_1^0\tilde{\chi}_1^0\gamma$ cross sections are penalized by a further power of $\alpha_{em}$ in the coupling.

Neutralinos are mixtures of the four fermionic SUSY partners of the neutral SM gauge and Higgs bosons [4]: the photino $\tilde{\gamma}$, the $Z$-ino $\tilde{Z}$ and the two higgsinos $\tilde{h}_a$ and $\tilde{h}_b$ (partners of the two Higgs-doublet neutral components). In the MSSM, all neutralino masses and couplings can be expressed at tree level in terms of three parameters [5]: $\mu$ (the SUSY Higgs-mixing mass), $M_2$ (the SU(2) gaugino mass, which also fixes all the other gaugino masses if unification conditions are imposed at the GUT scale, through, e.g., $M_1 = \frac{5}{3}\tan^2\theta_W M_2$), and $\tan\beta$ (the ratio of vacuum expectation values for the two Higgs doublets).[3] In order to study $e^+e^- \to \tilde{\chi}_1^0\tilde{\chi}_1^0\gamma$ cross sections, two further parameters are needed, i.e. the masses of the left and right selectrons, $m(\tilde{e}_{L,R})$. Since the corresponding contributions do not interfere, we assume them to be degenerate without loss of generality. Presently, the best direct experimental limit on the $\tilde{\chi}_1^0$ mass excludes the range $m_{\tilde{\chi}_1^0} < 20$ GeV for $\tan\beta > 2$. This limit disappears if $\tan\beta < 1.6$ [6].

---

[3]In this paper, we adopt the same notations as in [3].

Different physical components of neutralinos enter into different couplings with other particles. In particular, the gaugino ($\tilde{\gamma}$ and $\tilde{Z}$) components couple with fermions and their scalar SUSY partners, while the higgsino ($\tilde{h}_a$ and $\tilde{h}_b$) components couple to the $Z$. Hence, one can split the diagrams in Fig. 1 into two main subsets: a) three selectron-exchange ($t$-channel) diagrams, where only the $\tilde{\chi}_1^0$ gaugino part enters, and b) two $Z$-exchange ($s$-channel) diagrams, where only the higgsino $\tilde{\chi}_1^0$ components are relevant (assuming $m_e = 0$).

The limit $\tilde{\chi}_1^0 \equiv \tilde{\gamma}$ for the matrix element of the process $e^+e^- \to \tilde{\chi}_1^0\tilde{\chi}_1^0\gamma$ has been exactly evaluated in [7] (see also [8]). Here, we address the problem for the most general $\tilde{\chi}_1^0$ composition allowed in the MSSM, hence taking into account all the graphs shown in Fig. 1. Instead of evaluating the exact matrix element through a rather lengthy diagrammatic calculation, the spectrum of the observed photon is derived within the framework of the $p_t$-dependent structure-function approach [9], starting from the $e^+e^- \to \tilde{\chi}_1^0\tilde{\chi}_1^0$ cross section. This is a first application of the method of the so-called *single-photon library* recently discussed in the literature [10, 11]. By assessing its precision both in the $Z$-exchange and the selectron-exchange diagrams, the accuracy of the method is discussed and found to be within a few per-cent, and less than the foreseen experimental accuracy at LEP2. Higher-order QED corrections, due to undetected initial-state radiation, are also included. A comparison with the Standard Model main background from $e^+e^- \to \nu\bar{\nu}\gamma$ is performed for optimized photon kinematical cuts and quantitative conclusions on the signal/background ratio are given for a wide range of values of the SUSY parameters.

In order to get some initial hint on the dependence of the $e^+e^- \to \tilde{\chi}_1^0\tilde{\chi}_1^0\gamma$ cross section from the SUSY parameter space, we have preliminary looked at the behaviour of the cross section for the invisible process $e^+e^- \to \tilde{\chi}_1^0\tilde{\chi}_1^0$ in the plane $(\mu, M_2)$. This channel proceeds through either left and right selectrons, exchanged in the $t$-channel, or $Z$ vector bosons, exchanged in the $s$-channel. Analogously to the case of the $\tilde{\chi}_1^0\tilde{\chi}_2^0$ production, extensively discussed in [3], at $\sqrt{s} = 190$ GeV and for any value of $\tan\beta$, we can distinguish two asymptotic regime in the $(\mu, M_2)$ plane. For $|\mu| > M_Z$ and moderate values of $M_2$ and $m(\tilde{e}_{L,R})$, the dominant production mechanism is through selectron exchange, since the $\tilde{\chi}_1^0$ has dominant gaugino components. One can see that, for $m(\tilde{e}_{L,R}) = M_Z$, cross sections up to 1 pb and beyond are found in the gaugino regions not excluded by LEP1 data. Of course, these numbers are critically dependent on $m(\tilde{e}_{L,R})$. On the other hand, the higgsino's give the main contribution to the $\tilde{\chi}_1^0$ for $|\mu| < M_Z$ and $M_2 > M_Z$, hence enhancing the $Z$-channel production. Nevertheless, the particular combination of the higgsino components entering into the coupling with the $Z$ is quite depleted for a $\tilde{\chi}_1^0$ pair in the higgsino regions, especially at low values of $\tan\beta$ [3]. Hence, in these regions one expects a small $\sigma(\tilde{\chi}_1^0\tilde{\chi}_1^0)$ and, indeed, one can check that the cross section is always less than 0.1 pb in the area not excluded by LEP1 (not including QED corrections). This behaviour will reflect in a straightforward way on the $\sigma(e^+e^- \to \tilde{\chi}_1^0\tilde{\chi}_1^0\gamma)$ dependence in the plane $(\mu, M_2)$.

It is worth mentioning that an additional single-photon signature in the neutralino sector can arise at LEP2 from the $\tilde{\chi}_1^0\tilde{\chi}_2^0$ production followed by the one-loop radiative decay $\tilde{\chi}_2^0 \to \tilde{\chi}_1^0\gamma$ [12]. In particular parameter regions where the main $\tilde{\chi}_2^0 \to \tilde{\chi}_1^0 f\bar{f}$ decays are depleted, the branching ratio for the radiative $\tilde{\chi}_2^0$ decay can be of order 1 [13, 14]. A single-photon signal can also be associated to the $\tilde{\chi}_1^0\tilde{\chi}_2^0/\tilde{\chi}_2^0\tilde{\chi}_2^0$ production whenever the $\tilde{\chi}_2^0$ decays into invisible products ($\tilde{\chi}_2^0 \to \tilde{\chi}_1^0\nu\bar{\nu}$), through graphs analogous to those in Fig.1 with one or both $\tilde{\chi}_1^0$'s replaced by $\tilde{\chi}_2^0$'s. However, the latter signal will be in general depleted with respect to $e^+e^- \to \tilde{\chi}_1^0\tilde{\chi}_1^0\gamma$ by both the $\tilde{\chi}_2^0$ larger mass and its branching ratio into invisible products.

The outline of this paper is the following. In Section 2, we present the $p_t$-dependent structure-function approach that describes the observed photon associated to the process $e^+e^- \to \tilde{\chi}_1^0\tilde{\chi}_1^0\gamma$, as well as the formalism for implementing the higher-order QED corrections corresponding to undetected initial-state radiation. In Section 3, the photon energy distributions in some typical cases are discusses. The signal cross sections versus the main background from $e^+e^- \to \nu\bar{\nu}\gamma$ are studied in the SUSY parameter space, for optimized experimental cuts on the photon kinematical variables, in Section 4. In Section 5, we draw our conclusions.

## 2   The structure-function approach for radiative photons

At LEP2 energies, in view of the expected moderate experimental precision, simple approximate expressions can be used to compute cross sections and photon distributions for any process of the kind $e^+e^- \to \gamma + (invisible)$, where an undetectable final state is accompanied by the radiation of one hard large-angle photon.

In [10], a general approach is described that consists in dressing the exact tree-level cross section $\sigma_0^{e^+e^- \to (INV)}$ for the corresponding process $e^+e^- \to$

(*invisible*) with the angular radiator [15]

$$H^{(\alpha)}(x_\gamma, c_\gamma; s) = \frac{\alpha}{2\pi} \frac{1}{x_\gamma} \left[ 2 \frac{1+(1-x_\gamma)^2}{1+4m_e^2/s - c_\gamma^2} - x_\gamma^2 \right]. \tag{1}$$

$H^{(\alpha)}(x_\gamma, c_\gamma; s)$ describes the probability of radiating a photon with a given beam-energy fraction $x_\gamma = E_\gamma/E_b$ at the angle $\vartheta_\gamma$ ($c_\gamma \equiv \cos\vartheta_\gamma$). This radiator is derived from the $\mathcal{O}(\alpha)$ $p_t$-dependent structure function for the splitting $e \to e + \gamma$ [15]. It contains the leading infrared and collinear singularities plus a part of the next-to-leading contributions [given by the term $(-x_\gamma^2)$ in the squared brackets in eq. (1)], the latter providing a better description for hard photons at large angles. Then, the photonic spectrum for the reaction $e^+e^- \to \gamma + (invisible)$ is obtained by the following factorized expression

$$\frac{d\sigma}{dx_\gamma dc_\gamma} = \sigma_0^{e^+e^- \to (INV)}((1-x_\gamma)s) H^{(\alpha)}(x_\gamma, c_\gamma; s), \tag{2}$$

where the effective centre-of-mass energy for the bare cross section $\sigma_0^{e^+e^- \to (INV)}$ is reduced as a result of the photon radiation. This approximation amounts to attaching a photon line on the external electron legs, and to taking into account the "universal" part of the photon radiation with some of the next-to-leading effects.

The validity of this quite general method at LEP2 energies can be checked by applying it to the well-known Standard Model reaction $e^+e^- \to \nu\bar{\nu}\gamma$, for which the exact photon spectrum is known [16]. This process proceeds through two $s$-channel $Z$-exchange diagrams for all the three neutrino species, while for the electron neutrino case there are three further $t$-channel $W$-exchange diagrams. By inserting $\sigma_0^{e^+e^- \to \nu\bar{\nu}}$ as a kernel in the eq. (2) and comparing the result with the exact $e^+e^- \to \nu\bar{\nu}\gamma$ photon spectrum (including the three species of neutrinos), one finds a difference of about (1-3)% for $\sqrt{s} = (150-190)$ GeV (the error increasing with the energy) [10].

This check mainly tests the approximation for the $Z$-exchange channels, since these are common to all the neutrino species. Note also that the $W$-exchange diagram with the photon emitted by the internal $W$ propagator, contributing to the electron-neutrino channel, is not reproduced in this approximation. The agreement found shows that this contribution (that is not included in the "universal" factorization corresponding to eq. (2)) is not crucial at LEP2 energies. On the other hand, in the process $e^+e^- \to \tilde{\chi}_1^0 \tilde{\chi}_1^0 \gamma$, the $W$ exchanged in the $t$-channel is replaced by a scalar particle, i.e. a selectron. Because of the different spin, coupling and mass of the exchanged particle, one might wonder whether the effect of the non-universal part of the radiation (coming in this case from the photon emitted by the internal selectron propagator) is still under control. In order to settle this point, we have considered the limit $\tilde{\chi}_1^0 \equiv \tilde{\gamma}$ of the process $e^+e^- \to \tilde{\chi}_1^0\tilde{\chi}_1^0\gamma$. By doing so, one isolates the contribution from the selectron $t$-channels and can compare the results of the radiator approximation with the exact evaluation of the $e^+e^- \to \tilde{\gamma}\tilde{\gamma}\gamma$ spectrum [7]. At $\sqrt{s} = 190$ GeV and with typical cuts on the photon variables (cf. Section 4), the two results differ by less than 1.5% in the range of masses 0< $m(\tilde{\gamma})$ <75 GeV and 60 GeV< $m(\tilde{e}_{L,R})$ <100 GeV. The agreement gets better by increasing $m(\tilde{e}_{L,R})$, which corresponds to further depleting contributions from the "non-universal" radiation. Hence, by combining the accuracy for the $Z$-exchange graphs (assessed through the neutrino channel) and that for the selectron exchange, we can conclude that altogether the radiator-function approach in eq. (2) reproduces the exact $e^+e^- \to \tilde{\chi}_1^0\tilde{\chi}_1^0\gamma$ cross section with an accuracy better than a few per cent at LEP2.

In our treatment of the single-photon signal, we have also taken into account the higher-order QED corrections due to multiphoton soft emission and to the radiation of hard photons emitted in the very forward direction (and hence lost in the beam pipe), that give rise to undetected initial-state radiation. To this end, the spectrum obtained through eq. (2) has been convoluted with electron and positron structure functions [17]

$$\sigma(s) = \int dx_1\, dx_2\, dE_\gamma\, dc_\gamma\, D(x_1,s) D(x_2,s) \frac{d\sigma}{dE_\gamma dc_\gamma}, \qquad (3)$$

where $d\sigma/dE_\gamma dc_\gamma$ is the approximate spectrum for $e^+e^- \to \tilde{\chi}_1^0\tilde{\chi}_1^0\gamma$, the photon variables refer to the centre-of-mass frame after the initial-state radiation, and $D(x,s)$ is the electron (positron) structure function [18]. Eq. (3) is implemented in a Monte Carlo event generator based on [19], and used for the present analysis.

We stress again that the above procedure is quite general and can be applied whenever the contribution from "non-universal" photon radiation is not relevant.

## 3 Photon energy distributions

In this section, we study the energy distributions of the observed photon in $e^+e^- \to \tilde{\chi}_1^0\tilde{\chi}_1^0\gamma$, assuming as a typical LEP2 energy $\sqrt{s} = 190$ GeV. The results are obtained restricting the photon angle in the laboratory frame over the range $|c_\gamma| < 0.95$.

As discussed in the previous section, the accuracy of our radiator approach has been thoroughly tested in the higgsino sector of the process, mediated by the $Z$ exchange, through the channel $e^+e^- \to \nu\bar{\nu}\gamma$. On the other hand, we will see in the next section that observable rates for single-photon events from $e^+e^- \to \tilde{\chi}_1^0 \tilde{\chi}_1^0 \gamma$ can mainly be obtained in the gaugino sector, through the exchange of rather light selectrons. In order to assess the goodness of our radiator approach in the more exclusive case of the photon energy distribution, we start by analyzing the photon spectrum in the photino ($\tilde{\chi}_1^0 \to \tilde{\gamma}$) limit $e^+e^- \to \tilde{\gamma}\tilde{\gamma}\gamma$, for which the exact results are available. The photon energy spectra obtained in this case are given in Table 1 (for $m_{\tilde{\gamma}} = 20$ GeV), considering the contribution of only one selectron with mass $m(\tilde{e}) = 80$ GeV. The first column ($\Delta E_\gamma d\sigma_{ex}/dE_\gamma$) and the second one ($\Delta E_\gamma d\sigma_{app}/dE_\gamma$) show the results corresponding to the integration over different photon energy bins of the exact spectrum obtained through the exact matrix element in [7] and of the angular radiator approximation of eq. (2), respectively. The numerical integration error in the last one or two digits is shown in parenthesis. As can be seen, independently of the photino mass, the agreement between the exact and approximate spectrum is very satisfactory (well within 1%) for almost all the bins. It deteriorates at the level of some/several per cent quite near the kinematic limit $E_\gamma^{MAX}$, where non-universal radiation effects become non-negligible. Similar results have been obtained for $m_{\tilde{\gamma}} = 50$ GeV.

The excellent accuracy of the energy spectrum in the gaugino sector, when combined with the precision of our approximation in the $Z$-exchange (higgsino) channel, makes the radiator approach very reliable for the prediction of the photon spectrum for the most general $\tilde{\chi}_1^0$ composition.

In Fig. 2, we show the results obtained for the photon energy distributions of the process $e^+e^- \to \tilde{\chi}_1^0 \tilde{\chi}_1^0 \gamma$, assuming the following four scenarios

a) $\mu = -3M_Z, M_2 = M_Z, \tan\beta = 1.5$;

b) $\mu = 3M_Z, M_2 = M_Z, \tan\beta = 1.5$;

c) $\mu = -0.4M_Z, M_2 = 3M_Z, \tan\beta = 1.5$;

d) $\mu = 3M_Z, M_2 = M_Z, \tan\beta = 30$;

with $m(\tilde{e}_{L,R}) = M_Z$, at $\sqrt{s} = 190$ GeV and restricting the photon angle in the laboratory frame over the range $|c_\gamma| < 0.95$. The minimum photon energy considered is 1 GeV. In each plot (corresponding to one of the above scenarios), we compare the spectra with (solid lines) and without (dashed lines) the inclusion of higher-order QED effects.

In the first two cases [(a) and (b)], the gaugino components are dominant in the $\tilde{\chi}_1^0$ physical composition, and the spectra exhibit a typical $t$-channel structure, although a small bump at large energies in the case (b) reveals some $Z$ radiative-return effect, due to some non-vanishing higgsino component in the $\tilde{\chi}_1^0$. This effect is quite apparent in the case (c), where the $\tilde{\chi}_1^0$ is mainly a higgsino, and the dominance of the $s$-channel reflects into a strong peak near the maximum energy value. One consequence of the Z radiative return, is that the role of the initial state radiation in the modification of the shape becomes particularly significant. Notice also that the absolute normalization in the latter spectrum corresponds to quite smaller production rates with respect to the previous scenarios. The last case (d) just shows the effect of increasing the value of $\tan\beta$ in the case (b).

## 4   Signal versus the $e^+e^- \to \nu\bar{\nu}\gamma$ background

In this section, we present results for the single-photon total rates for the process $e^+e^- \to \tilde{\chi}_1^0\tilde{\chi}_1^0\gamma$ in the SUSY parameter space $(\mu, M_2)$, for both low and high values of $\tan\beta$. The cross sections are obtained by applying eq. (2) and also higher-order QED effects arising from soft and collinear (undetected) photons.

All our results, are obtained by integrating the angular photon variable over the range $|c_\gamma| < 0.95$ (corresponding to about $\vartheta_\gamma > 18°$) and by imposing a cut on the minimum transverse momentum of the photon

$$(p_T)_\gamma > 0.065 E_b,$$

where $E_b$ is the beam energy. The latter is the typical cut needed is order to avoid veto angles in the LEP2 experimental apparatus, where the background from radiative Bhabha-scattering events could fake the signal [11].

A last cut is imposed on the maximum photon energy. This helps to optimize the rejection of the main background arising from $e^+e^- \to \nu\bar{\nu}\gamma$. Indeed, due to the Z radiative-return effect, after applying the above cuts, the neutrino $E_\gamma$ spectrum is mostly concentrated at large energies, in particular around the value $(1 - M_Z^2/s)\sqrt{s}/2$ [10]. By imposing the further cut

$$E_\gamma < 0.5 E_b,$$

one reduces the neutrino cross section to about one third of the original value. In particular, at $\sqrt{s} = 190$ GeV one finally gets $\sigma_{cut}(\nu\bar{\nu}\gamma) \simeq 1.4$ pb (including higher-order QED effects). On the other hand, as far as the signal is concerned, the latter cut mainly penalizes the higgsino sector of the $\tilde{\chi}_1^0$, where a large fraction of the events is associated to the Z radiative-return peak (cf. Section 3). Nevertheless, production rates in this sector of the $(\mu, M_2)$ plane turn out to be in general small, even before applying the cut $E_\gamma < 0.5 E_b$, in regions not excluded by LEP1.

The neutralino cross sections obtained applying this set of cuts are shown in the plane $(\mu, M_2)$ in Fig. 3, for $\tan\beta = 1.5$, and in Fig. 4, for $\tan\beta =$

30. Considering the canonical LEP2 integrated luminosity of 500 pb$^{-1}$, for $m(\tilde{e}_{L,R}) = M_Z$, cross sections up to about 30 fb are found in the gaugino regions, corresponding to up to 15 events. This signal has to fight against a neutrino background of about 700 events, with a corresponding $S/\sqrt{B}$ ratio of at most 0.57. On the other hand, one gets smaller rates in the higgsino regions, that is at most O(10) fb in the area not excluded by LEP1 data.

The main effect of including higher-order QED is an enhancement of the $s$-channel $Z$-exchange contributions, due to the shifting of part of the photon energy spectrum from the radiative-return peak (which is cut off in our treatment) down to lower energy values (cf. Section 3). In the higgsino regions, where the $Z$-exchange diagrams are relevant, a considerable increase of the rates is observed. On the other hand, in the more interesting gaugino regions the cross sections slightly decrease (cf. Fig. 2).

In Fig. 5, the effect of lowering the selectron masses down to 60 GeV is shown, at small $\tan\beta$. While no relevant change is of course observed in the higgsino regions, the gaugino cross sections increase by a factor 1.7-2, with a corresponding $S/\sqrt{B}$ ratio up to about 1. On the contrary, heavier selectron masses can give rise to drastically lower neutralino rates. For instance, in Fig. 6, the case of $m(\tilde{e}_{L,R}) = 3M_Z$ is considered, at small $\tan\beta$. Hence, an observable single-photon rate could arise mainly from selectrons light enough to be pair produced at LEP2.

# 5 Conclusions

The production of invisible pairs of lightest neutralinos in association with a large-angle hard photon at LEP2 energies has been analysed in the MSSM. For the first time, a comprehensive study is made assuming the *most general* gaugino/higgsino composition of the $\tilde{\chi}_1^0$. Instead of relying upon an explicit diagrammatic calculation, we have calculated the spectrum of the observed photon in the process $e^+e^- \to \tilde{\chi}_1^0\tilde{\chi}_1^0\gamma$ through a $p_t$-dependent structure function approach. The accuracy of the method is assessed to be within the foreseen experimental LEP2 precision. The photon energy distributions as well as the total cross sections for $e^+e^- \to \tilde{\chi}_1^0\tilde{\chi}_1^0\gamma$ as functions of the parameters $(\mu, M_2)$ have been discussed, for both low and high values of $\tan\beta$. In order to provide reliable predictions, the results include the effects of higher-order QED corrections and take into account realistic experimental photon cuts for the LEP2 data analysis.

It turns out that the presence of the irreducible Standard Model background arising from $e^+e^- \to \nu\bar{\nu}\gamma$ makes the observation of a SUSY signal from $e^+e^- \to \tilde{\chi}_1^0\tilde{\chi}_1^0\gamma$ quite difficult at LEP2, even in the most favourable regions of the parameter space. In the $(\mu, M_2)$ regions where the lightest neutralino is mainly a higgsino, the cross sections, although not affected by the selectron mass uncertainties, are too small to be measurable. In the remaining parameter space, both photinos and $Z$-ino components give moderate rates for scalar masses not much larger than the LEP2 exclusion limit $[m(\tilde{e}_{L,R}) \simeq M_Z]$. Hence, a clear single-photon signal from neutralinos at LEP2 would only arise together with a direct selectron pair production.

The present study shows a first application beyond the Standard Model of the idea of building up a *single-photon library*. Starting from any process with invisible final states, by following the same procedure described here for the process $e^+e^- \to \tilde{\chi}_1^0 \tilde{\chi}_1^0 \gamma$, one can reconstruct in a straightforward way the corresponding single-photon signal through our $p_t$-dependent structure function approach. Further applications of this method to different processes giving rise to the single-photon signature will be discussed in forthcoming papers.

### Acknowledgements

We would like to thank Graham Wilson for useful discussions.

# References

[1] H. E. Haber and G. L. Kane, Phys. Rep. 117 (1985) 75.

[2] A. Bartl, H. Fraas and W. Majerotto, Nucl. Phys. B278 (1986) 1.

[3] S. Ambrosanio and B. Mele, Phys. Rev. D52 (1995) 3900.

[4] J. Ellis and G.G. Ross, Phys. Lett. 117B (1982) 397; J. M. Frère and G. L. Kane, Nucl. Phys. B223 (1983) 331; J. F. Gunion and H. E. Haber, Nucl. Phys. B272 (1986) 1; B402 (1993) 567.

[5] A. Bartl, H. Fraas, W. Majerotto and N. Oshimo, Phys. Rev. D40 (1989) 1594.

[6] L3 Collaboration, M. Acciarri et al., Phys. Lett. 350B (1995) 109.

[7] K. Grassie and P. N. Pandita, Phys. Rev. D30 (1984) 22; J. Ware and M. Machacek, Phys. Lett. 142B (1984) 300; see also M. Chen, C. Dionisi, M. Martinez and X. Tata, Phys. Rep. 159 (1988) 201.

[8] P. Fayet, Phys. Lett. 117B (1982) 460; J. Ellis and J.S. Hagelin, Phys. Lett. 122B (1983) 303.

[9] O. Nicrosini and L. Trentadue, Phys. Lett. B231 (1989) 487, and references therein.

[10] G. Montagna, O. Nicrosini, F. Piccinini and L. Trentadue, Nucl. Phys. B452 (1995) 161.

[11] F. Boudjema, B. Mele et al., "Standard Model Processes", in *Physics at LEP2*, G. Altarelli, T. Sjostrand and F. Zwirner eds., Report CERN 96-01, vol. 1, p. 207 (1996), hep-ph/9601224.

[12] H. Komatsu and J. Kubo, Phys. Lett. **157B** (1985) 90; Nucl. Phys. **B263** (1986) 265; H. E. Haber, G. L. Kane and M. Quirós, Phys. Lett. **160B** (1985) 297; Nucl. Phys. **B273** (1986) 333; G. Gamberini, Z. Phys. **C30** (1986) 605; H. E. Haber and D. Wyler, Nucl. Phys. **B323** (1989) 267.

[13] S. Ambrosanio and B. Mele, Phys. Rev. D53 (1996) 2541.

[14] S. Ambrosanio and B. Mele, "Supersymmetric Scenarios with Dominant Radiative Neutralino decay", Preprint ROME1-1148/96 (1996).

[15] O. Nicrosini and L. Trentadue, Nucl. Phys. B318 (1989) 1, and references therein.

[16] F. A. Berends et al., Nucl. Phys. B301 (1988) 583.

[17] G. Montagna, O. Nicrosini and F. Piccinini, Phys. Rev. D48 (1993) 1021, and references therein.

[18] E. A. Kuraev and V. S. Fadin, Sov. J. Nucl. Phys. 41 (1985) 466;
G. Altarelli and G. Martinelli, *Physics at LEP*, CERN Report 86–02, J. Ellis and R. Peccei, eds. (Geneva, 1986); see also:
O. Nicrosini and L. Trentadue, Phys. Lett. B196 (1987) 551; Z. Phys. C39 (1988) 479.
For a review see also:
O. Nicrosini and L. Trentadue, in *Radiative Corrections for $e^+e^-$ Collisions*, J. H. Kühn, ed. (Springer, Berlin, 1989), p. 25; in *QED Structure Functions*, G. Bonvicini, ed., AIP Conf. Proc. No. 201 (AIP, New York, 1990), p. 12; O. Nicrosini, ibid., p. 73.

[19] G. Montagna, O. Nicrosini and F. Piccinini, "NUNUGPV - A Monte Carlo event generator for $e^+e^- \to \nu\bar{\nu}\gamma(\gamma)$ events at LEP", FNT/T-96/1, in press on Computer Physics Communications.

| $\Delta E_\gamma (GeV)$ | $\Delta E_\gamma d\sigma_{ex}/dE_\gamma (fb)$ | $\Delta E_\gamma d\sigma_{app}/dE_\gamma (fb)$ |
|---|---|---|
| 1 - 2 | 9.624(4) | 9.633(11) |
| 2 - 3 | 5.596(2) | 5.601(6) |
| 3 - 4 | 3.946(2) | 3.950(5) |
| 4 - 5 | 3.043(1) | 3.045(3) |
| 5 - 6 | 2.471(1) | 2.473(3) |
| 6 - 7 | 2.0769(8) | 2.079(2) |
| 7 - 8 | 1.7883(7) | 1.790(2) |
| 8 - 9 | 1.5680(6) | 1.569(2) |
| 9 - 10 | 1.3941(6) | 1.395(2) |
| 10 - 15 | 5.273(2) | 5.275(6) |
| 15 - 20 | 3.627(1) | 3.627(4) |
| 20 - 25 | 2.726(1) | 2.724(3) |
| 25 - 30 | 2.1566(8) | 2.153(2) |
| 30 - 35 | 1.7635(7) | 1.759(2) |
| 35 - 40 | 1.4752(6) | 1.470(2) |
| 40 - 45 | 1.2537(5) | 1.249(1) |
| 45 - 50 | 1.0771(4) | 1.073(1) |
| 50 - 55 | 0.9313(4) | 0.927(1) |
| 55 - 60 | 0.8070(3) | 0.804(1) |
| 60 - 65 | 0.6969(3) | 0.694(1) |
| 65 - 70 | 0.5955(2) | 0.593(1) |
| 70 - 75 | 0.4974(2) | 0.493(1) |
| 75 - 80 | 0.3966(1) | 0.3866(5) |
| 80 - 85 | 0.2859(1) | 0.2614(4) |
| 85 - 90.8 | 0.1589(1) | 0.1046(2) |

Table 1: Photon energy distributions for $e^+e^- \to \tilde{\gamma}\tilde{\gamma}\gamma$ (limit: $\tilde{\chi}_1^0 \to \tilde{\gamma}$) for $m_{\tilde{\gamma}} = 20$ GeV and $m(\tilde{e}_L) = 80$ GeV. Exact (first column) versus approximated (second column) results at $\sqrt{s} = 190$ GeV and with $|\cos\vartheta_\gamma| < 0.95$. The contribution of only one scalar electron ($\tilde{e}_L$) is considered.

# Figure Captions

Figure 1. Feynman diagrams for the process $e^+e^- \to \tilde{\chi}_1^0 \tilde{\chi}_1^0 \gamma$.

Figure 2. Photon energy distributions in $e^+e^- \to \tilde{\chi}_1^0 \tilde{\chi}_1^0 \gamma$ in four representative cases (a), (b), (c) and (d). The selectron masses are set at $m(\tilde{e}_{L,R}) = M_Z$ and $\sqrt{s} = 190$ GeV. The $\tilde{\chi}_1^0$ mass corresponding to each choice of SUSY parameters is also shown. The spectra with (solid lines) and without (dashed lines) the inclusion of higher-order QED effects are compared.

Figure 3. Cross sections for $e^+e^- \to \tilde{\chi}_1^0 \tilde{\chi}_1^0 \gamma$ at small $\tan\beta$, including higher-order QED corrections.

Figure 4. Cross sections for $e^+e^- \to \tilde{\chi}_1^0 \tilde{\chi}_1^0 \gamma$ at large $\tan\beta$, including higher-order QED corrections.

Figure 5. Cross sections for $e^+e^- \to \tilde{\chi}_1^0 \tilde{\chi}_1^0 \gamma$ at small $\tan\beta$, including higher-order QED corrections. Effect of decreasing the selectron mass.

Figure 6. Cross sections for $e^+e^- \to \tilde{\chi}_1^0 \tilde{\chi}_1^0 \gamma$ at small $\tan\beta$, including higher-order QED corrections. Effect of increasing the selectron mass.

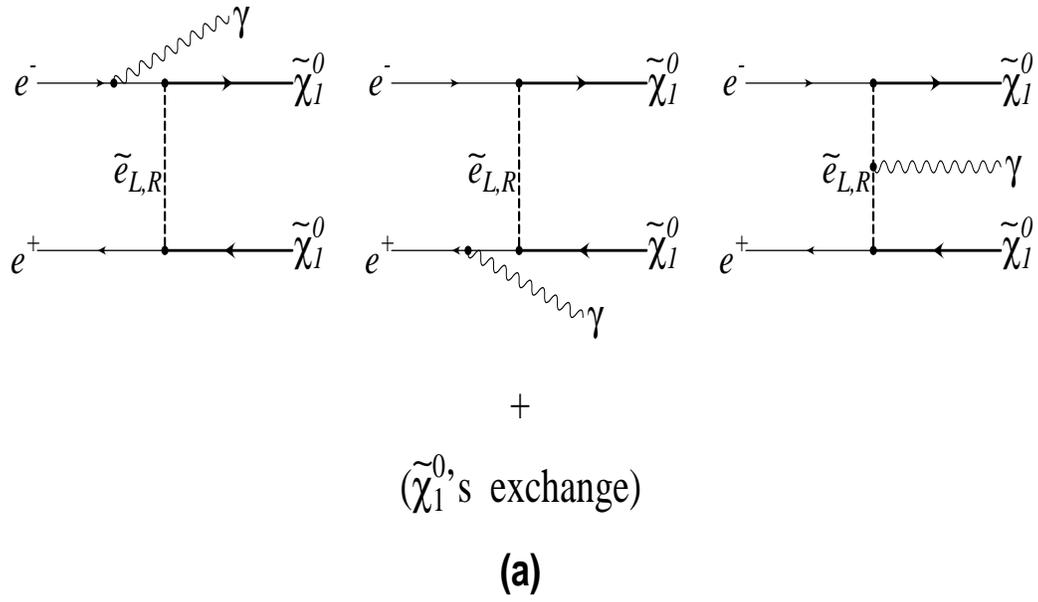

+

($\tilde{\chi}_1^0$'s exchange)

**(a)**

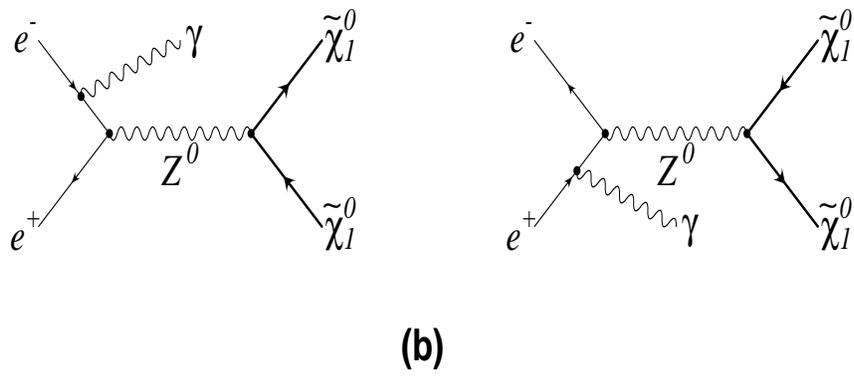

**(b)**

Figure 1:

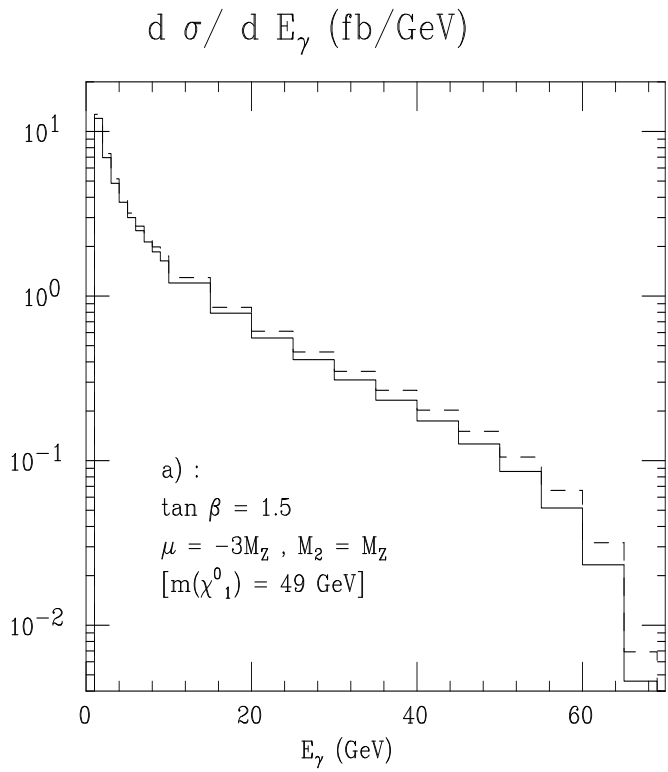
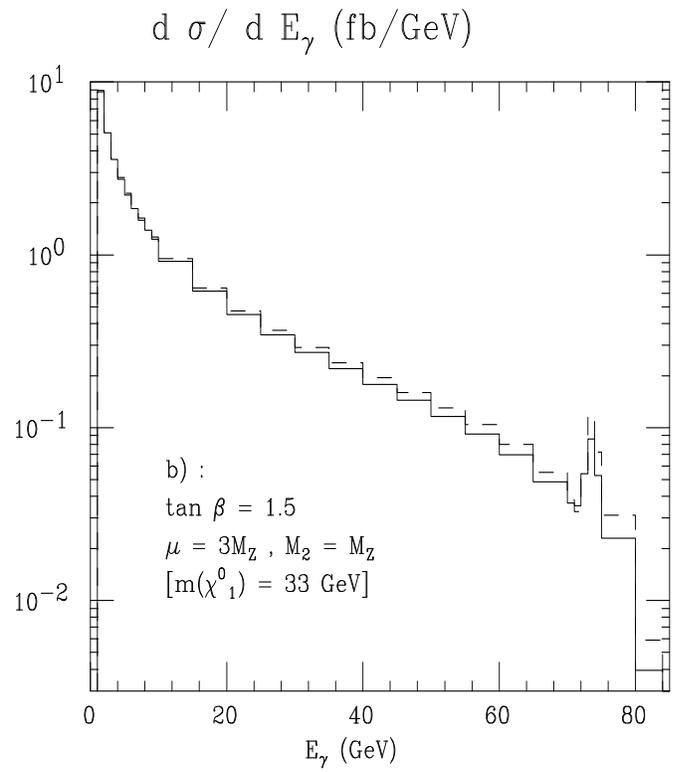
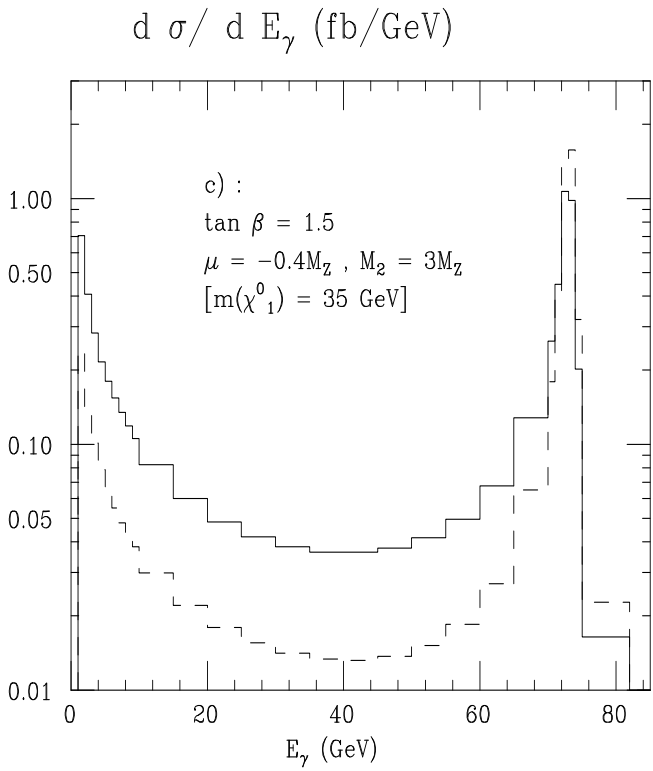
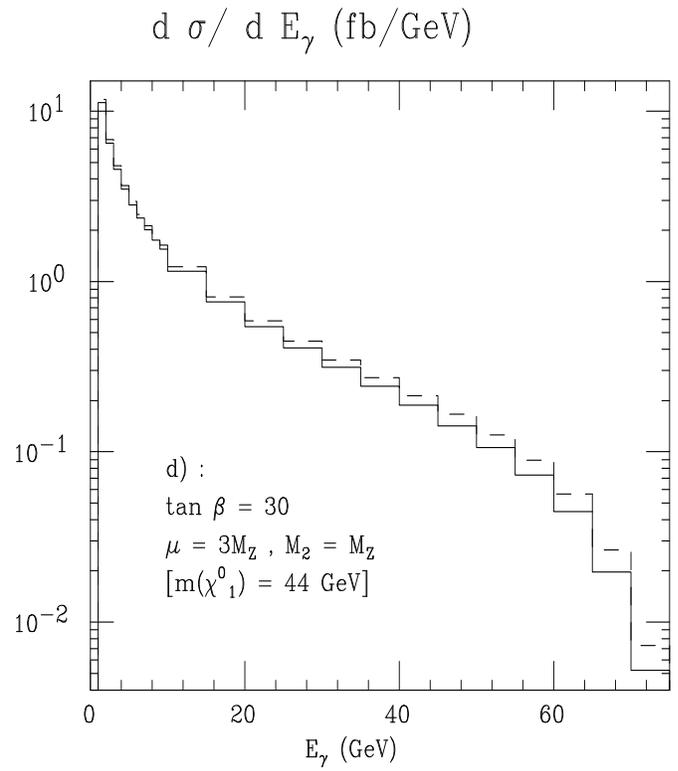

Figure 2:

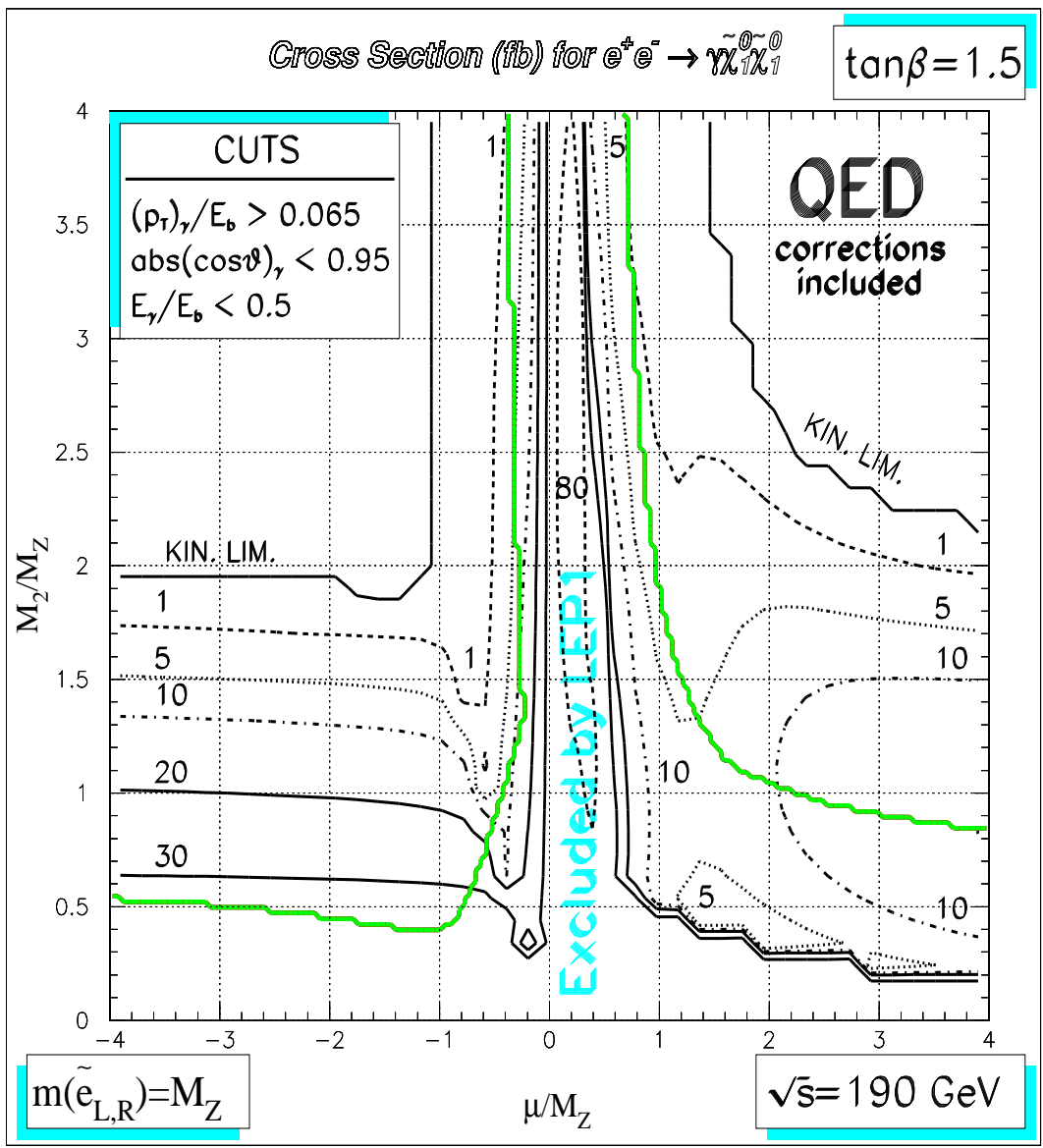

Figure 3:

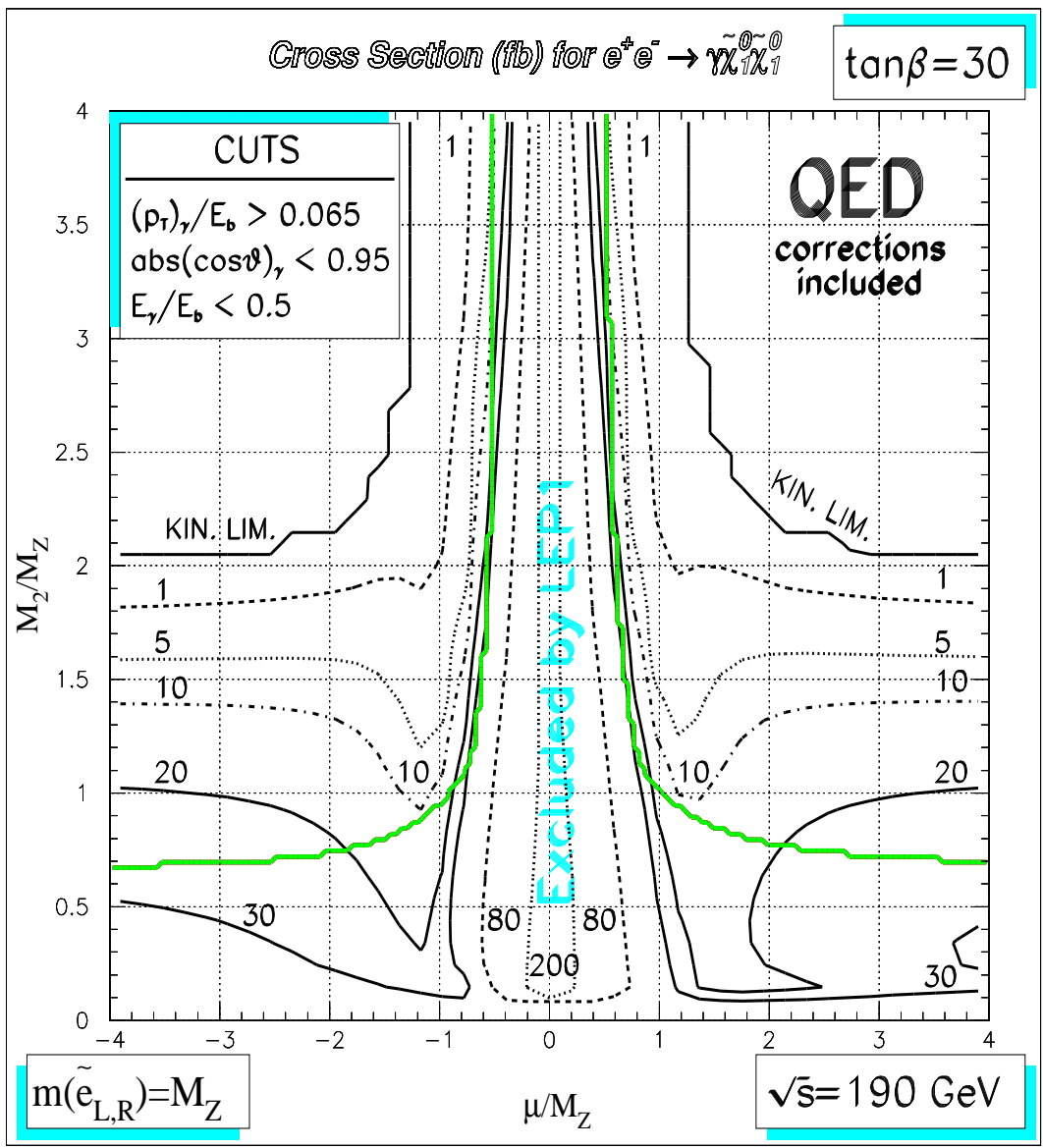

Figure 4:

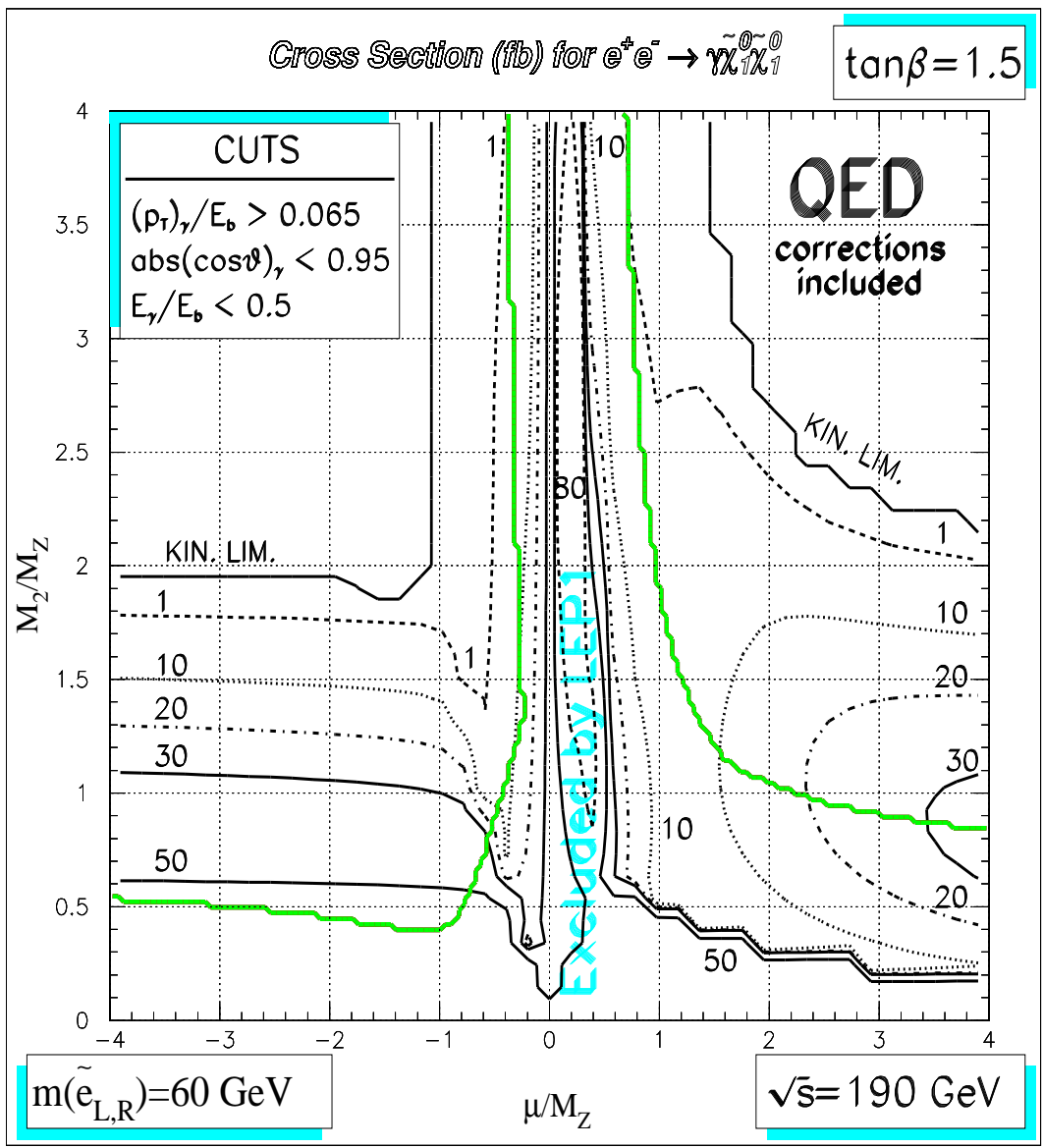

Figure 5:

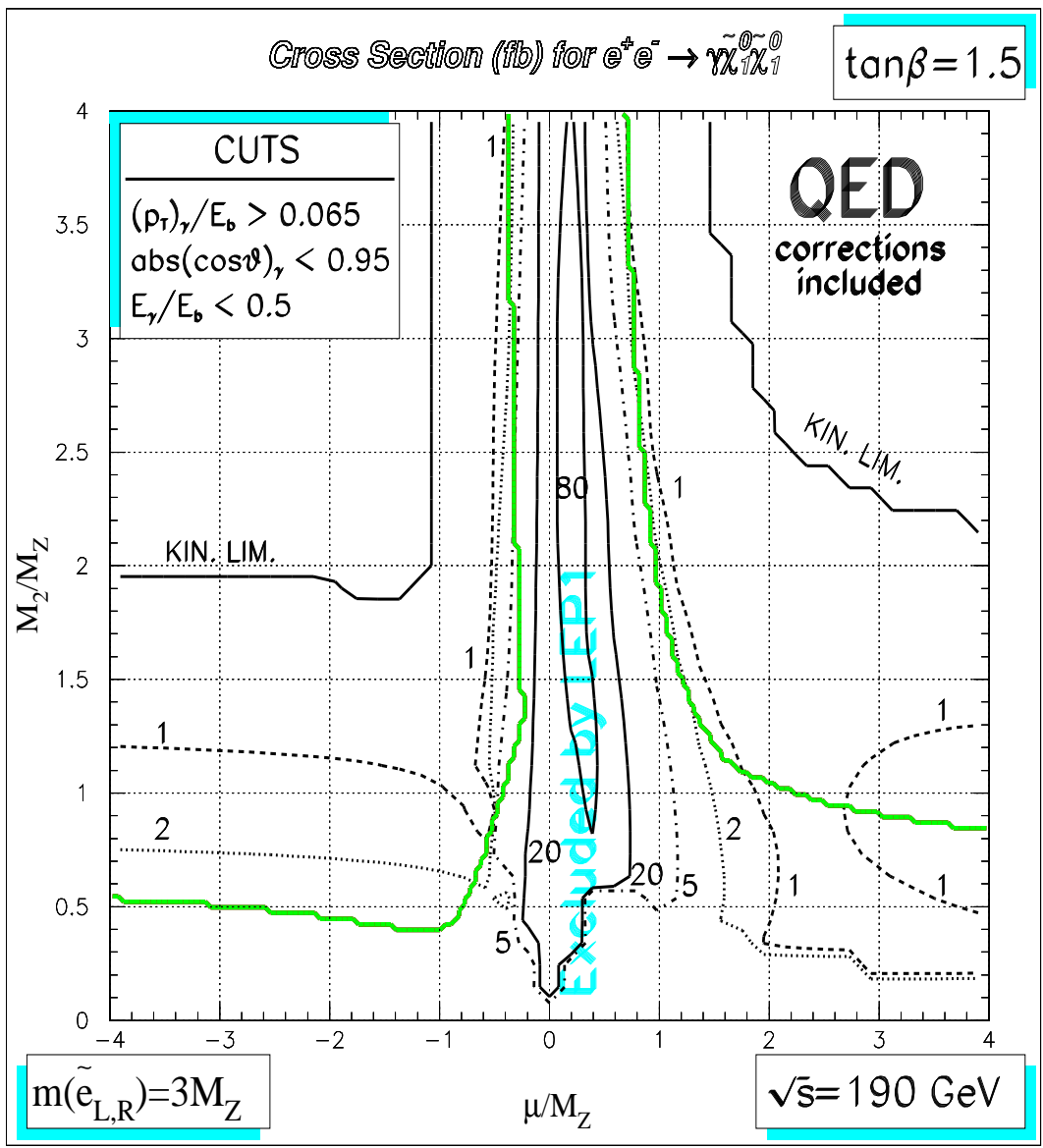

Figure 6: